\documentclass[graphicx,3p,sort&compress,plb]{elsarticle}
\usepackage{lscape}
\usepackage{longtable}
\usepackage{latexsym,amsbsy,amsmath,amsbsy}
\newcommand{\be}{\begin{equation}}
\newcommand{\ee}{\end{equation}}
\newcommand{\SpSp}[2]{ \mbox{$\vec{\sigma }_{#1}.\vec{\sigma }_{#2}$}}
\newcommand{\lala}[2]{ \mbox{${\tilde{\lambda}_{#1}\cdot
         \tilde{\lambda}_{#2}}$}}
\renewcommand{\vec}[1]{\boldsymbol #1}
\newcommand{\df}[2]{\ensuremath{ {\raise
1pt\hbox{$\displaystyle #1$}\over \raise -2pt \hbox{$\displaystyle
#2$}}}}
 \makeatletter
 \newcommand*{\rom}[1]{\expandafter\@slowromancap\romannumeral #1@}
 \makeatother
\begin{document}
\begin{frontmatter}
\title{\bf Isovector and hidden-beauty partners of  the \boldmath $X(3872)$ \unboldmath} 
\date{\today}
\author[oslo]{Hallstein~H\o gaasen}
\ead{hallstein.hogasen@fys.uio.no}
\address[oslo]{Department of Physics, University of Oslo,
Box 1048  NO-0316 Oslo Norway}
\author[orsay]{Emi Kou}
\ead{kou@lal.in2p3.fr}
\address[orsay]{Laboratoire de l'Acc\'elérateur Lin\'eaire, Universit\'e Paris-Sud, IN2P3-CNRS\\[-1pt]
Centre Scientifique d'Orsay,
91898 Orsay Cedex, France }
\author[lyon]{Jean-Marc~Richard}
\ead{j-m.richard@ipnl.in2p3.fr}
\address[lyon]{Universit\'e de Lyon, Institut de Physique Nucl\'eaire de Lyon, 
 UCBL--IN2P3-CNRS\\[-1pt]
4, rue Enrico Fermi, 69622 Villeurbanne cedex, France}
\author[annecy]{Paul~Sorba}
\ead{paul.sorba@lapth.cnrs.fr}
\address[annecy]{LAPTh, Laboratoire d'Annecy-le-Vieux de Physique Th\'eorique\\[-1pt]
CNRS, Universit\'e de Savoie, BP\,110, 74941 Annecy-le-Vieux Cedex, France }
%
%
\begin{abstract}
The isovector partners of the $X(3872)$, recently found at BES\,\rom{3}, Belle 
 and CLEO-c were predicted in a simple model based on the chromomagnetic interaction among quarks . The extension to the hidden-beauty sector is discussed.
\end{abstract}


\end{frontmatter}

\section{Introduction}\label{se:int}
Recently, a new hidden-charm meson was seen at BES\,III and Belle \cite{Ablikim:2013mio,Liu:2013dau}. Its remarkable feature, as compared to most previous $X$, $Y$, $Z$ states is that it carries an electric charge.  It is currently named $X(3900)^+$. Shortly after its announcement, its existence was confirmed by the Northwestern group working on CLEO-c data \cite{Xiao:2013iha}, who also have some indication for the neutral member of the isospin triplet. 

Note that three other charged states with hidden charm have been observed, $Z(4050)^+$, $Z(4250)^+$ and $Z(4450)^+$, but only by the Belle collaboration.
The $Z(4050)^+$ and $Z(4250)^+$ have been seen by Belle in the $B$ decay \cite{Mizuk:2008me}, but not confirmed in a search by Babar \cite{Lees:2011ik}. 
The $Z(4450)^\pm$ was seen by Belle in the $\pi^\pm\psi'$ invariant mass of the $B\to K \pi^\pm\psi'$ decay \cite{Choi:2007wga,Mizuk:2009da}, and the quantum numbers $1^+$ are favoured \cite{Chilikin:2013tch}. To our knowledge, this state was not confirmed in other channels or other experiments. 

Two charged states have been seen in the hidden-beauty sector, the $Z_b(10610)^\pm$ and the $Z_b(10650)^\pm$, again by the Belle collaboration \cite{Belle:2011aa}. The latest result deals with the $Z_b(10610)^0$ discovered by Belle \cite{Krokovny:2013mgx}, the neutral partner of the $Z_b(10610)^\pm$.

The $X(3872)$ has $J^{PC}=1^{++}$ as early indicated in several experiments (see, e.g.,  \cite{Beringer:1900zz}), and confirmed recently at the  Large Hadron Collider of CERN (LHC)(see, e.g., the analysis by LHCb \cite{Manca:2013zma}). 
The simplest scenario is that the new $X(3900)^+$ has the same~$J^{P}$ quantum numbers as the $X(3872)$, namely $J^{P}=1^{+}$.  

A major issue is whether the $X$, $Y$ and $Z$ states are mostly molecules, i.e., bound states or resonances made of a flavoured meson and an anti-flavoured meson, or mostly a tetraquark states in which the quark interact directly. An analysis of the production rate of X(3872) in \cite{Bignamini:2009sk,Bignamini:2009fn} indicates that the measured cross section at Tevatron is
too large for a molecule interpretation, even after taking into account the re-scattering effect suggested in \cite{Artoisenet:2009wk}.
%

The problem is to find a simple explanation for the approximate degeneracy of the isospin $I=0$ and $I=1$ states. In the molecular model, the $X(3872)$ is mainly a $D\bar{D}{}^*+\mathrm{c.c.}$ state, and an important contribution to binding comes from the one-pion exchange, which includes an isospin-dependent factor $\vec\tau_1.\vec\tau_2$ whose absolute value is weaker for $I=1$ than $I=0$. \footnote{There is also a change of sign for 
$\vec\tau_1.\vec\tau_2$, which is $+$ for $I=1$ and $-$ for $I=0$, but the pion-mediated interaction is off-diagonal in the $\{D\bar D{}^*,D^*\bar D\}$ basis, and thus the attractive or repulsive character depend on which of the $D\bar D{}^*\pm D^*\bar D$ combination is considered.}
\@
In short, the molecular model of $X,Y,Z$ states favours isospin $I=0$ states, as did earlier the nucleon--antinucleon model of the baryonium resonances \cite{Buck:1977rt}.

On the other hand, the quark model with a flavour-independent interaction gives a natural explanation to ``exchange-degeneracy'', with, e.g., $\omega$ and $\rho$ exactly degenerate as long as the quark-antiquark internal annihilation and the coupling to decay channels are neglected.
Thus if the $X(3872)$ and the $X(3900)^+$ have the same~$J^P$, it is tempting to seek an explanation in terms of quark dynamics, rather than in a molecular picture.\footnote{Of course, in case of identical quarks, the Pauli principle can induce some isospin dependence from the spin dependence. This is the reason why the $\Lambda$ baryon is lighter than the $\Sigma$ one. But here, this effect is not present, as isospin is carried by a quark and an antiquark.}\@ Indeed, some models based on quark dynamics have predicted the isospin $I=1$ state $X(3900)^+$ near its $I=0$ partner$X(3872)$. This is the case for the chromomagnetic model discussed below and for the diquark model of Ref.~\cite{Faccini:2013lda}.

This property of exchange degeneracy illustrates the similarities and differences between QED and QCD.  After the work of  De R\'ujula, Giorgi and Glashow \cite{DeRujula:1975ge}, it has become widely accepted that the pattern of spin-spin splittings in quark models is similar in structure to that of the hyperfine splittings in atomic physics, namely is due to an interaction among 
chromomagnetic moments. In the case of the positronium atom, the interaction between the magnetic moments explains only about half of the energy difference between the spin-triplet and the spin-singlet states. The hyperfine splitting in positronium receives a substantial contribution from the annihilation diagram where the electron-positron pair goes into a single virtual photon and back to an electron-positron pair. 
For the usual quark-antiquark mesons, there is no such effect, as the gluon transforms as an octet in colour. But the effect can show up for multiquarks, in which a quark-antiquark pair can be in a colour octet state. We shall discuss later the role of the Pirenne potential, when suitably adapted from QED to quark models.

The aim of this article is to revisit how the isospin $I=1$ partner of the $X(3872)$ was predicted in a simple model \cite{Hogaasen:2005jv}, and to discuss to which extent the model can be extended toward the hidden-beauty sector. 
\boldmath \section{The $X(3872)$ and the $X(3900)^+$ in a chromomagnetic model}\unboldmath
\label{se:X}
Some years ago, three of us proposed a simple model for the $X(3872)$ state, described as a $(c\bar cq \bar q)$ tetraquark \cite{Hogaasen:2005jv}, in which both the $(c\bar c)$ pair and the $(q\bar q)$ pair are mostly in a colour-octet state. This structure prevents the state from dissociating freely into a charmonium and a light meson. More precisely, the dynamics of the 
$(c\bar cq \bar q)$ in \cite{Hogaasen:2005jv} is governed by the chromomagnetic Hamiltonian 
\begin{equation}\label{eq:Hcm}
H=M+H_{\mathrm{CM}} =\sum_i m_i - \sum_{i,j} C_{ij}\,\lala{i}{j}\,\SpSp{i}{j}~,
\end{equation}
where the $m_i$ are effective quark masses including the chromo-electric effects, 
and $\tilde\lambda_i$ and $\vec\sigma_i$ the colour and spin operator acting on the $i^{\rm th}$ quark, with suitable changes for an antiquark. Should one start from an explicit potential model, then $\sum_i m_i$ would stand from the expectation value of the mass and kinetic-energy term, and the last term in \eqref{eq:Hcm} represents the expectation value of the spin--spin interaction. Thus $C_{ij}$ includes the intrinsic strength of the chromomagnetic potential divided by the quark masses, and multiplied by the short-range 
correlation of the quarks $i$ and $j$. In principle, these terms should vary from a ground state hadron to another one. An empirical observation is that the quantities $m_i$ and $C_{ij}$ are nearly constants for $i$ or $j$ denoting $u,\,d,\,s$ or $c$, suggesting the possibility of extrapolating from simple to more complicated configurations. A good surprise in our attempt \cite{Hogaasen:2005jv} is that one of the eigenstates of \eqref{eq:Hcm} has some of the key properties of  $X(3872)$. 

Moreover, Ref.~\cite{Hogaasen:2005jv} contains a \emph{prediction} for the isospin $I=1$ partner of $X(3872)$, at 3900\,MeV. In the discussion following Eq.\ (10) of \cite{Hogaasen:2005jv}, it is stated that ``the mostly $I=1$ state lies 31\,MeV above the mostly $I=0$ state.''
This calculation includes a mixing effect, as the quark masses $m_u$ and $m_d$ are taken to be different.

In the neutral sector, the $I=0$ and $I=1$ states are left degenerate by the chromomagnetic Hamiltonian \eqref{eq:Hcm}. Introducing the contribution of the annihilation diagram and different masses for the $u$ and $d$ quark give an additional contribution in the  $\{(c\bar c u\bar u),\,(c\bar c d\bar d)\}$ basis which reads
\begin{equation}
 \label{eq:Pirenne}
\delta H=\begin{pmatrix}
 2\,m_u-a & -a \\
-a & 2\,m_d -a
\end{pmatrix}~.
\end{equation}

We now have to fix the value of the parameter $a$ governing the annihilation term. In the positronium atom, the virtual process $e^+ + e^-\to\gamma\to e^+ + e^-$ contributes to the hyperfine splitting, in addition to the Breit-Fermi interaction. The effect is given by the Pirenne potential~\cite{Pirenne:1947}. Its strength is three times that of the Breit-Fermi contact interaction. 
The analogue for QCD has been discussed in the context of studies on baryonium and other exotic states \cite{Hogaasen:1979dq,Chan:1980cj,Gelmini:1979ir}. In the perturbative limit, there is an additional factor 2 due to colour, besides the factor 3 in QED. However, as stressed by Gelmini, the annihilation is substantially  suppressed by the confinement of the gluons. So, instead of $a=6\, C_{q\bar q}$, a choice $a\sim C_{q\bar q}$ is reasonable.

In \cite{Hogaasen:2005jv}, the values 
$a=15\,$MeV and $m_d-m_u=3.5\,$MeV were adopted,  leading to a difference of about $31\,$MeV between the two eigenvalues, leading the the prediction of about 3904\,MeV for the neutral $I=1$ partner of the $X(3872)$. 

For the charged states of the $I=1$ multiplet, $\delta H$ is simply replaced by $m_u+m_d$, and this puts the charged states about
$0.4\;$MeV below the neutral, mostly isovector,  one.

\section{Extension to the hidden-beauty sector}
\label{se:hidden-b}
The difficulty is our model \eqref{eq:Hcm} consists of identifying a single effective mass for a flavoured quark in open-flavour mesons, flavoured baryons and hidden-flavour mesons. The combinations
\begin{equation}
 \label{eq:masses}
\begin{aligned}
3\,(Q\bar q)_{S=1}+(Q\bar q)_{S=0}&=4\,m_Q+4\,m_q~,\\
2\,\Sigma_Q^*+\Sigma_Q+\Lambda_Q&=4\,m_Q+8\,m_q~,\\
3\,(Q\bar Q)_{S=1}+(Q\bar Q)_{S=0}&=8\,m_Q~,
\end{aligned}
\end{equation}
should be compatible, and in particular, one should verify
\begin{equation}
\label{eq:mass-relation}
\delta M=12\,(Q\bar q)_{S=1}+4\,(Q\bar q)_{S=0}-4\,\Sigma_Q^*
{}-2\,\Sigma_Q-2\,\Lambda_Q-3\,(Q\bar Q)_{S=1}-(Q\bar Q)_{S=0}=0~.
\end{equation}
In the charm sector, one gets $\delta M\simeq -200\,$MeV, which is rather satisfactory, but for the beauty sector, the results is $\delta M\simeq 1000\,$MeV. It indicates that the bottomonium states gives an average quark mass $m_b=4721\,$MeV, much lighter than the combination $m_b=(12\,B^*+4\,B-4\,\Sigma_b^*-2\Sigma_b-2\,\Lambda_b)/8=4852\,$MeV deduced from heavy-light systems. This is due to the strong chromoelectric attraction between two heavy quarks in $(b\bar b)$.

We thus generalize our model to include a chromoelectric term, and replace \eqref{eq:Hcm} by
\begin{equation}
\label{eq:Hcem}
H =M+H_\text{CE}+H_\text{CM}
=\sum_i m_i- \sum_{i,j} A_{ij}\,\lala{i}{j}-\sum_{i,j} C_{ij}\,\lala{i}{j}\,\SpSp{i}{j}~.
\end{equation}

Introducing a few non-vanishing chromo-electric coefficients $A_{ij}$ implies a change of the effective masses. 
A minimal solution is found with $m_q=450\,$MeV, $m_c=1530\,$MeV, $m_b=4860\,$MeV, and all $A_{ij}=0$, except for $A_{bb}=53\,$MeV by fitting the spin-averaged ground-state masses of $(c\bar c)$, $(c \bar q)$, $(cqq)$ and the $c\to b$ analogues. A slightly better agreement is found by allowing both $A_{cc}$ or $A_{bb}$ to be non-zero, but we shall keep the minimal solution. 

We use the basis defined in \cite{Hogaasen:2005jv}, namely
\begin{equation}\label{eq:alphai}
\begin{aligned}
\alpha_1 &=(q_1 \overline{q}_3)^1_0 \otimes (q_2
\overline{q}_4 )^{{1}}_{1},\quad 
\alpha_2 =(q_1 \overline{q}_3)^1_1 \otimes (q_2 \overline{q}_4
)^{{1}}_{0}~,\\
\alpha_3 &=(q_1 \overline{q}_3)^1_1 \otimes (q_2
\overline{q}_4 )^{{1}}_{1},\quad 
\alpha_4 =(q_1 \overline{q}_3)^8_0 \otimes (q_2 \overline{q}_4
)^{{8}}_{1}~, \\ 
\alpha_5 &=(q_1 \overline{q}_3)^8_1 \otimes (q_2
\overline{q}_4 )^{{8}}_{0}, \quad
\alpha_6 =(q_1 \overline{q}_3)^8_1 \otimes  (q_2 \overline{q}_4
)^{{8}}_{1}~.
\end{aligned}
\end{equation}
where the superscript denotes the colour 1 or 8, and the subscript 0 or 1 denotes the spin, with an overall recoupling to 
a colour-singlet $J^P=1^+$ state.

The matrix elements of the colour-magnetic part have been given in \cite{Hogaasen:2005jv}, and are reminded in Table~\ref{tab:cm} for completeness.
\newpage
\begin{landscape}
\begin{table}[c]
\caption{\label{tab:cm}Colourmagnetic Hamiltonian $-H_\mathrm{CM}$ in
the basis (\ref{eq:alphai})}
\vbox{%
$$\ \ \left[\vbox{\hskip -5pt
\begin{tabular}{cccccc}
$16C_{13}-\df{16}{3}C_{24}$&0&0&0&$\df{8\sqrt2}{3}(C_{23}+C_{12})$&0\\
0&$-
\df{16}{3}C_{13}+16C_{24}$&0&$\df{8\sqrt2}{3}(C_{23}+C_{12})$&0&0\\
0&0&$-\df{16}{3}(C_{13}+C_{24})$&0&0&$\df{8\sqrt2}{3}(C_{23}-C_{12})$\\
0&$\df{8\sqrt2}{3}(C_{23}+C_{12})$&0&$\df{2}{3}C_{24}-2C_{13}$&
$\df{28}{3}C_{23}-\df{8}{3}C_{12}$&0\\
$\df{8\sqrt2}{3}(C_{23}+C_{12})$&0
&0&$\df{28}{3}C_{23}-\df{8}{3}C_{12}$&
$-2 C_{24}+ \df{2}{3}C_{13}$&0\\
0&0&$\df{8\sqrt2}{3}(C_{23}-C_{12})$
&0&0&$\df{2}{3}(4C_{12}+14C_{23}+C_{13}+C_{24})$
\end{tabular}}\ \right]$$
}
\end{table}
\end{landscape}

One should now supplement it by the matrix elements of the chromo-electric term, which are
\begin{equation}\label{eq:ce}
H_\text{CE}=
\begin{pmatrix}
X_a & 0 & 0 & X_b& 0 & 0 \\
 0 &X_a & 0 & 0 & X_b & 0 \\
 0 & 0 &X_a & 0 & 0 &X_b\\
 X_b & 0 & 0 &X_c & 0 & 0 \\
 0 & X_b & 0 & 0 & X_c & 0 \\
 0 & 0 & X_b & 0 & 0 &X_c\\
\end{pmatrix}~,
\end{equation}
with 
\begin{align}\label{eq:ce1}
X_a&= -\frac{16}{3}\,(A_{13}+A_{24})~,\nonumber\\
X_b&=\frac{4\,\sqrt{2}}{3}\,(A_{12}+A_{34}-A_{14}-A_{23})~,   \\
X_c&= \frac{2\,(A_{13}+A_{24}) -4\,(A_{12}+A_{34})-14\,(A_{14}+A_{23})}{3}~. \nonumber    
      \end{align}

The parameters are summarized in Table~\ref{tab:para}.
\begin{table}[!htb]
 \caption{\label{tab:para} Parameters of the model: masses $m_i$, non-vanishing chromoelectric $A_{ij}$ and chromomagnetic $C_{ij}$ coefficients (in MeV)}
\begin{center}
 \begin{tabular}{ccccccccc}
$m_q$ & $m_c$ & $m_b$  & $A_{bb}$ & $C_{qq}$ & $C_{qc}$ & $C_{cc}$ & $C_{qb}$ & $C_{bb}$ \\
450.  & 1530. & 4860.  & 52.  & 20. & 6. & 5.2 & 1.9 & 3.2 
\end{tabular}
\end{center}
\end{table}

The ground-state masses of heavy quarkonia and heavy light mesons obtained using these parameters are listed in Table~\ref{tab:ordinary}

\begin{table}[!htb]
 \caption{\label{tab:ordinary} Masses or mass differences of ground state hadrons in the model (in GeV)}
\begin{center}
\begin{tabular}{lccccccc}
State & $J/\psi$ & $J/\psi-\eta_c$ & $D$ & $D^*-D$ & $\Lambda_c$ & $\Sigma_c-\Lambda_c$ & $\Sigma^*_c-\Sigma_c$\\
Exp. &  3.10 & 0.117 & 1.87 & 0.141 &  2.29 & 0.166 & 0.065\\
Model & 3.09 & 0.111 & 1.88 & 0.128 &  2.27 & 0.149 &  0.096
\end{tabular}
\vskip 9pt
\begin{tabular}{lccccccc}
State & $\Upsilon$  & $\Upsilon-\eta_b$ & $B$  & $B^*-B$ &$\Lambda_b$ & $\Sigma_b-\Lambda_b$ & $\Sigma^*_b-\Sigma_b$\\
Exp. &   9.46 & 0.069 & 5.28 & 0.046&  5.62 & 0.194 & 0.020\\
Model & 9.46 & 0.068 & 5.28 & 0.041& 5.60 & 0.193 &  0.030
\end{tabular}
\end{center}
\end{table}
\section{Results}
The Hamiltonian is now diagonalized, using the parameters of Table~\ref{tab:para} fitting some ground-state ordinary hadrons containing the same quarks, $q$, $c$, $b$ and the associated antiquarks.

In the $(c\bar c q\bar q)$ sector, one obtains results identical to the ones reported in \cite{Hogaasen:2005jv}, with in particular, a
state of mass very close to 3872\,MeV which is a pure $\alpha_6$ state. It was then argued that if $C_{\bar qc}$ is taken slightly larger than $C_{qc}$, then a small $\alpha_3$ component is admixed, that is responsible for the observed decay of $X(3872)$ into $J/\psi$ and a light vector meson. In our model, when
the wave function  is expressed in the $(c\bar q)(\bar c q)$ basis, it has a  large
colour-singlet--colour-singlet components which corresponds to a decay
into $D\bar D{}^*$ or c.c., which is, however, strongly suppressed by the lack
of phase-space for the $X(3872)$.

The $X(3900)^+$ is less known experimentally. We  refer to a very recent review by Olsen
\cite{Olsen:2014mea}\footnote{This paper was posted after the first version of
the present article}. The width is given as $46\pm22\,$MeV. The decay proceeds mainly
through $D\bar D{}^*+\text{c.c.}$, and benefits for this channel from a much more favourable
phase-space than for the X(3872) \cite{Ablikim:2013xfr}.
Our model predicted a dominance of this decay into a charm-carrying
vector plus a charmed pseudoscalar configuration when phase-space opens up. In contrast to what happens for the $X(3872)$, this superallowed decay becomes more important than the decays into $(c\bar c)+(q\bar q)$.

In this latter sector, the discovery channel of the $X(3900)^+$ was  $J/\psi + \pi$. In our model, as in the case of the $X(3872)$, introducing $C_{\bar q c}
\neq C_{qc}$ generates a small $\alpha_3$ component in the wave function
of the $X(3900)^+$ that induces a decay into $J/\psi$ and a charged vector meson.
The $J/\psi+\pi$ decay involves a $\alpha_2$ component that is not provided in our simple model. Similarly, a  decay involving $\eta_c$ would require a $\alpha_1$ component, or a spin-flip in the decay, which is suppressed, as discussed, e.g., in \cite{Voloshin:2004mh}.

In the hidden-beauty sector, one gets an analogue state of mass about 10.62\,GeV, and a wave function $\sum_i b_i\,\alpha_i$ with $\{b_i\}\propto\{0, 0, 0,0,0,1\}$. This means that this is a pure octet-octet state, so that the fall-apart decay into $(b\bar b)+(q\bar q)$ is suppressed. This state is about 11\,MeV above the $B\bar B{}^*$ threshold, and thus slightly more unstable with respect to this threshold, as compared to the $X(3872)$ with respect to the $D\bar D{}^*$ threshold. As for the $X(3872)$, introducing some departure from $C_{bq}=C_{b\bar q}$ would induce a small component consisting of a $J/\psi$ and a light vector meson. 

As the breaking of exchange degeneracy and isospin symmetry occurs through  light quarks, we except the same spacings between isospin $I=0$ and $I=1$ as in the hidden-charm sector, and same spacing among the neutral and charged states in the $I=1$ triplet. 

Note that for the quartet of $(b\bar b q\bar q)$ states predicted  near 10.62\,GeV, the chromoelectric term gives a repulsion of about 35\,MeV. As, e.g., when deriving the short-range part of the nucleon-nucleon interaction~\cite{Harvey:1980rva}, estimating the masses and properties of multiquark states implies some speculation on the colour dependence of the effective interaction. The chromoelectric term in Eq.~\eqref{eq:Hcem} corresponds to a colour-octet exchange, which is the most reasonable choice for a pairwise interaction, as a colour-singlet exchange would confine everything together. But multi-body forces could be envisaged in more complicated models.
\section{Summary and conclusions}
In this article, it was reminded that a simple quark model predicted the existence of a $I=1$ partner of the $X(3872)$ at the right mass and thus anticipated the recent discovery by BES\,III, Belle and CLEO-c~\cite{Ablikim:2013mio,Liu:2013dau,Xiao:2013iha}. The model consisted of effective masses and a chromomagnetic interaction. It can be supplemented by a minimal chromoelectric term and then applied to the sector of states with hidden-beauty.

The model predicts a nearly degenerate quartet (an $I=0$ singlet and an $I=1$ triplet, with some mixing of the neutrals) near 10.62\,MeV. The charged states are possible candidates for either the $Z_b(10610)^\pm$ or $Z_b(10650)^\pm$ states of Belle \cite{Mizuk:2008me}. It is, however, very difficult in this approach to produce an isospin $I=1$ state without a nearby $I=0$ partner, and to arrange two nearly degenerate isotriplets. 

It seems important to use the most advanced accelerators and detectors to investigate this sector of hadron physics. The Belle\,\rom{2} facility \cite{Akeroyd:2004mj} will of course provide us with crucial information. But the search is already active at the LHC, with in particular, a very recent  search for the $X_b$ by the CMS collaboration \cite{Chatrchyan:2013mea}, with no evidence in the $\Upsilon(1S)\pi^+\pi^-$ channel. It is hoped that the combined efforts at lepton and hadron colliders will definitely clarify the situation in the hidden-beauty sector.
\section*{Acknowledgements}
We thank B.\ Grinstein for a useful correspondence. 

\newpage

\end{document}